# Rare-earth Doped Amorphous Silicon Microdisk and Microstadium Resonators with Emission at 1550nm

D.S.L. Figueira and N. C. Frateschi

Instituto de Física "Gleb Wataghin", Universidade Estadual de Campinas, UNICAMP 13083-970, Campinas, São Paulo, Brazil and Centro de Componentes Semicondutores, Universidade Estadual de Campinas, UNICAMP 13083-970, Campinas, São Paulo, Brazil

**Abstract**

Microdisks and microstadium resonators were fabricated on erbium doped amorphous hydrogenated silicon (a-Si:H<Er>) layers sandwiched in air and native $SiO_2$ on Si substrates. Annealing condition is optimized to allow large emission at 1550 nm for samples with erbium concentrations as high as $1.02 \times 10^{20}$ atoms/cm$^3$. Near field scanning optical microscopy shows evidences of the simultaneous presence of bow-tie and diamond scars. These modes indicate the high quality of the resonators and the potentiality for achieving amorphous silicon microcavity lasers.

**Indexing terms**: Silicon photonics, microdisk, microstadium, directional emission.

# I. INTRODUCTION

The development of silicon based active optical devices has been of interest in recent years due the possibility of integration with CMOS technology. Silicon doping with rare-earth elements is an alternative for efficient light emission in the C-band.

Microdisk resonators with whispering gallery modes (WGM) [1], offer great advantage for stimulated light emission generation in small volumes where long photon lifetime is achieved with simple processing steps [1-2]. Also, light emission occurs mainly along the substrate plane being potentially suitable for photonic integration. Moreover, with the use of non-conventional structures such as stadium, one may achieve higher emission spectral and directional control, expanding the integration possibilities [3].

Rare-earth elements inserted in some semiconductors and insulators emit luminescence due to internal transitions of the incomplete *4f* levels in wavelengths that are almost independent of the matrix. The *f-f* transition is forbidden, but it is partially allowed when the wave functions are mixed to those of opposing parity. This always occurs on localized ions in non-central symmetrical lattices [4]. The PL process demands that the *4f* electrons are excited absorbing energy. In materials with large band gap, as in glasses, this occurs in general with a direct absorption to one of the transitions of upper energy of the rare-earth ion [5]. In semiconductors, the excitation process has a more complicated mechanism [6-7]. Particularly, in hydrogenated amorphous silicon doped with erbium (a-Si:H<Er>), it appears that transfer of energy from the matrix to $Er^{3+}$ ions occurs [7-8] because the excitation spectrum follows the absorption spectra of the the a-Si:H. Therefore, continuous pumping with a wavelength that is absorbed by the a-Si:H is possible. It has been proposed that the excitation mechanism of Er in a-Si:H is a defect related Auger resonant process between the $Er^{3+}$ $^4I_{15/2}$ and $^4I_{13/2}$ states and the dipole formed by the s-like conduction band states and p-like dangling bond states [6].

Therefore, combining the high resonator quality factor microdisk structures with $Er^{3+}$ doped Si material is a natural approach to achieving new photonic components compatible with Si technology.

In this work we present our development of microdisks and microstadium resonators based a-Si:H<Er> sandwiched in air and $SiO_2$ obtained by wet oxidation of Si substrates as in CMOS process. Photoluminescence is optimized by annealing time and temperature to allow large emission in the C-band. Finally, we map the near field behavior of the resonator emission.

## II. SAMPLE PREPARATION, CHARACTERIZATION, AND MICROSTRUCTURES FABRICATION TECHNIQUES

Initially we prepared a film of 1.25μm in thickness of $SiO_2$ in Si (type "n" <100>) substrates by wet oxidation at a temperature of $1100^oC$ for 270 minutes in a flux of 1.0 l/min of $O_2$ and water vapor. Ellipsometry was performed at 632.8 nm for the measurement of $SiO_2$ refraction index and thickness with a value of 1.46 and 1250 nm, respectively. A refractive index of 1.44 at 1550 nm is obtained by interferometric techniques using a thin film analyzer and the thickness value agrees with profilometry results. The a-Si:H<Er> films were prepared by RF sputtering co-deposition on $Si/SiO_2$ substrates. The base pressure of the vacuum chamber was $2\times10^6$ mbar and the sputtering was carried out in RF mode with the bias fixed at 1kV from circular sources in an atmosphere of $8\times10^{-3}$ mbar of argon and a flux of 8.0 sccm of hydrogen for the Si passivation [9]. As source, we used crystalline Si wafer mixed with metallic erbium. The substrate temperature was maintained at 240°C and the deposition rate of roughly 0.5Å/s. This technique is a direct form to obtain simultaneously a-Si:H film doped with $Er^{3+}$ unlike other reported methods requiring plasma deposition followed by Er implantation [10].

The refractive index of the as deposited material was measured by ellipsometry at 1550 nm and interferometry with values of 3.84 and 4.1, respectively. The extinction factor was found to be 0.198. The discrepancies between the two techniques are within the measurement error. The film was 350 nm in thickness. This is a high index value very suitable for light confinement between a-Si:H<Er> and air. For this thickness, only one vertical mode is allowed. The effective index of refraction for the TE and TM modes are 3.4 and 2.9, respectively in a slab waveguide immersed in air for λ=1550 nm. Confinement above 90% for both TE and TM modes is expected. The calculated penetration depth (1/e) of the confined mode in the pedestal area is smaller than 50 nm for both TE and TM modes allowing the use of thin oxide layers with reduced optical mode leakage.

Using rutherford backscattering spectrometry we have obtained 0.36% of Er, 96.00% of Si, 3.54% of Ar and 0.06% of Cu. An Er concentration of $1.02 \times 10^{20}$ atoms/cm$^3$ is achieved. This value, which can easily be incremented with co-deposition, is already above the typical concentration obtained by ion implantation [10-12]. The copper contamination comes from the substrate holder. The copper contamination is within the resolution of the measurement. The argon presence is expected in this type of growth where it is common to find trapped argon in the film [13].

After co-deposition, photoluminescence (PL) was used to investigate the best annealing condition for the optimization of emission efficiency. PL was performed using the excitation of on the argon laser line at 514nm and 200mW with a spot of approximately 100μm of radius at room temperature with an incidence angle of approximately 30° from the normal. PL emission, perpendicular from the sample plane, was focused at the entrance slit of a 2m spectrometer and a liquid nitrogen cooled Ge photodetector was used at the exit slit to obtain the spectrum. Both slits were kept at 200μm. The sample was annealed at a temperature between 0°C and 500°C for 30min in a quartz tube oven in $N_2$ atmosphere at 3.0 l/min flux. Fig. 1(a) shows the PL emission intensity dependence on annealing temperature. Clearly, an optimum condition is near 300 °C. The inset in figure 1 shows the typical PL spectrum obtained for the samples. All samples presented emission at 1550nm at room temperature. The second observed PL peak in figure 1 is caused by the split of atomic levels caused by Stark effect [14]. Annealing time was also investigated. Fig. 1(b) shows the dependence of PL emission intensity with annealing time. The optimum condition is near 30 min. Indeed annealing is important in enhancing the PL efficiency of these films resulting in an enhancement of over 2.5 times.

Fabrication of the microdisk and microstadium resonator involved basically three steps, beginning with photolithography to define the resonator shape with AZ5214 photoresist by UV exposure at approximately 400 nm. Disks of 5 μm radius and stadia composed of two semi-circles with a radius of R=15 μm adjacent to a rectangle with sides $L_1$= R and $L_2$ = 2R were fabricated. The eccentricity of these stadia, defined by $\varepsilon = \dfrac{L_2/2}{L_2/2 + 2R}$, is 1/3. $SF_6/CF_4/CHF_3$ (10/10/3 sccm) plasma dry etch was optimized to transfer the resist etch mask into the a-Si:H<Er> with good morphology. The plasma etch was realized in a home-made reactive ion eating (RIE) system, under 50mTorr of pressure and 50W RF power. This was followed by a buffer HF (BOE) wet etch to remove $SiO_2$ for the definition of pedestal. The microdisks were supported by a small diameter $SiO_2$ pillar. A SEM image of a typical processed microdisk is shown in the inset of fig. 2.

## III. LIGHT EMISSION CHARACTERIZATION AND DISCUSSION

Figure 2(a) shows the PL spectrum of a stadium pumped by an Ar-ion laser at 514 nm focused to a spot with a diameter of approximately 40 μm. Besides the smaller beam diameter, the set-up is identical to the PL measurement described above and the measured emission if off the disk plane. An unprocessed sample was measured under the same condition for comparison, figure 2(b). A 2x reduction in emission intensity is observed for the processed stadium. Given the stadium dimension, about 50% of the pump beam is off the structure. Therefore, processing is not leading to any considerable damage.

Mapping the light emission of the microstadium was realized by Topometrix near-field scanning optical microscopy (NSOM). With a Nd:Yag doubled at 532nm pumping the structures, it was possible to measure NIR emission using a 1550 nm photomultiplier tube (PMT-NIR Hamamatsu H10330) with a 532 nm blocking filter. The difference in wavelength between this laser and the one used for PL is not expected to be relevant given the indirect pumping mechanism of the a-Si:H<Er>. All measurements were performed at room temperature. The structures were pumped perpendicularly by a tapered fiber with a 100 nm tip and the 1550 nm emission was collected by the photomultiplier tube at approximately 30º from the normal. The circular microdisks presented isotropic emission. Fig. 3 shows the measured near-field for the microstadium. Higher light emission along the major axis is observed. Also, higher optical density is observed at the four corners of the stadium. The stadium cavity drawn with the emission image is a guide to the eye allowing only a qualitative analysis.

Periodic orbits, called scars, [15-16] were predicted for electrons confined in 2D stadium boundaries by Bunimovich [17]. Analogously, for stadium microdisk resonators, it is expected that the optical density would occur preferentially at these scars [18-19]. The measured emission of our microstadium apparently shows enhanced emission occurring in two typical scars: bow-tie and diamond. These two scars were added to figure 3 to help visualization. The bow-tie mode is confined within the cavity and the observed light is essentially generated by scattering at the reflection points. On the other hand, the diamond scar results with large emission at the curved reflection points. No emission was observed on the straight wall reflection points. Also, the presence of low Q Fabry-Perot modes between the two curved walls may be enhancing the emission along the major axis. The coexistence of the two scars is expected since they are the orbits with the smallest round trip time for scars with large photonic lifetime. The large photonic lifetime of these modes is achieved in a range of eccentricity for which the incidence angle at the reflection points is larger than the critical angle for quasi-total internal reflection [20]. Figure 4 shows the incidence angle dependence on stadium eccentricity for both scars. A dashed line shows the critical angle for total internal reflection calculated for the effective TE index. Both diamond and bow-tie scars are expected to have incidence angle above total internal reflection and, thus, high photonic lifetime, in a range of $0.151 < \varepsilon < 0.53$. The eccentricity of our stadium was chosen such that it lies approximately in the middle of this range. The presence of these modes indicates the high quality of the resonators. This may be a first step for the fabrication of amorphous silicon microcavity lasers.

**CONCLUSION**

We have presented a direct process for fabrication films of Si/SiO2/ a-Si:H<Er> with applications in the manufacture of structures for silicon photonics compatible with conventional CMOS process. $Er^{+3}$ concentrations in excess of $1.02 \times 10^{20}/cm^3$ are obtained by RF sputtering co-deposition. A strong PL emission at 1550 nm is observed. Annealing of the samples shows a strong dependence of the PL emission with annealing temperature. Very good morphology microdisk and microstadium resonators were fabricated with emission near 1550 nm. Near field scanning optical microscopy shows isotropic emission from disks and the coexistence of bow-tie and diamond scars. The presence of these modes indicates the high quality of the resonators and the possibility of achieving amorphous silicon microcavity lasers.

**Acknowledgments:** The authors would like to thank Francisco Paulo M. Rouxinol for assistance in RBS analysis and Wendel L. Moreira for help in Near-Field Scanning Optical Microscope (NSOM) measurements. This work was supported by the Conselho Nacional de Pesquisa (CNPq) and the Fundação de Amparo à Pesquisa do Estado de São Paulo (FAPESP).

**Figure captions :**

**Fig. 1.** Photoluminescence intensity under different annealing conditions. Bottom axis: (a) annealing temperature. Top axis: (b) annealing time; Maximum PL intensity is around 300ºC with an annealing time of 30min. The inset show a typical PL spectra obtained in a-Si:H<Er> samples at room temperature .

**Fig. 2**. Small spot photoluminescence spectrum at room temperature of (a) Unprocessed a-Si:H<Er> film. (b) Microstadium. The inset shows a SEM micrograph of a 5μm radius a-Si:H<Er> microdisk. The disk is supported by a $SiO_2$ pedestal 1.25μm tall with 3.2μm diameter.

**Fig. 3.** Near-field scanning optical microscopy (NSOM) mapping of light emission for the microstadium. The diamond and bow-tie scars are drawn to help visualization.

**Fig.4.** Calculated dependence of incidence angle with stadium eccentricity for bow-tie (solid triangles) and diamond scars (open triangles). The dashed line shows the critical incidence angle for total internal reflection.

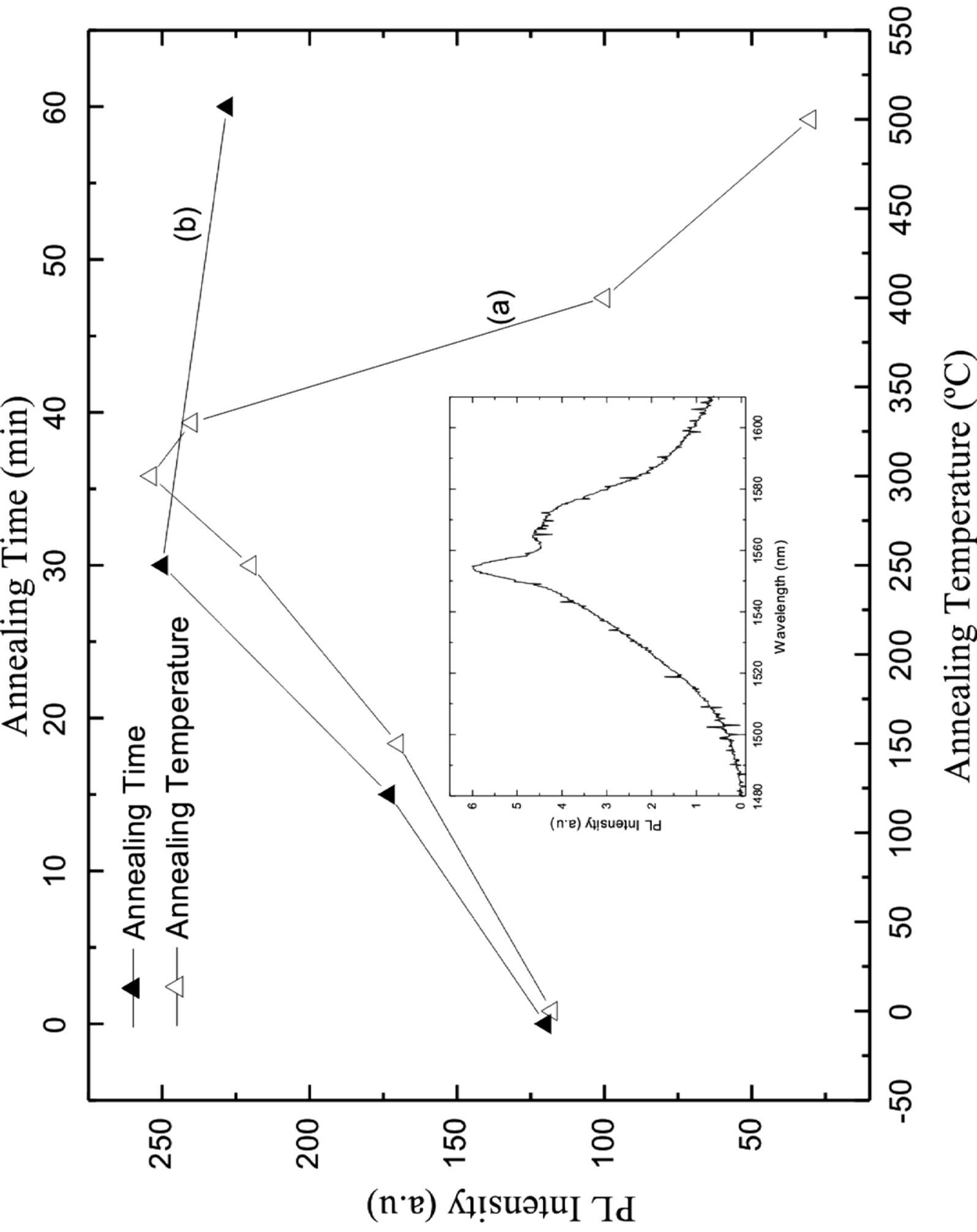

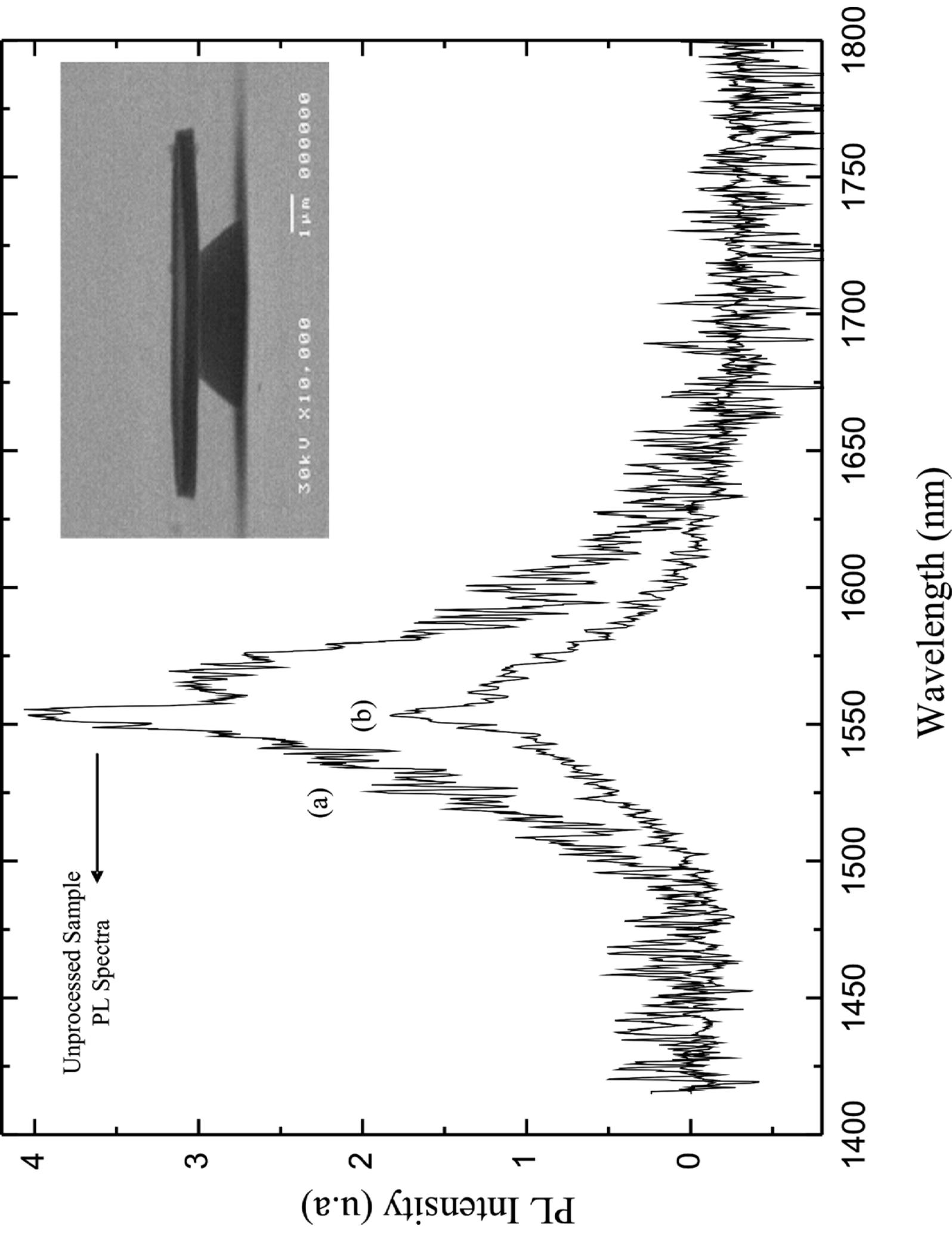

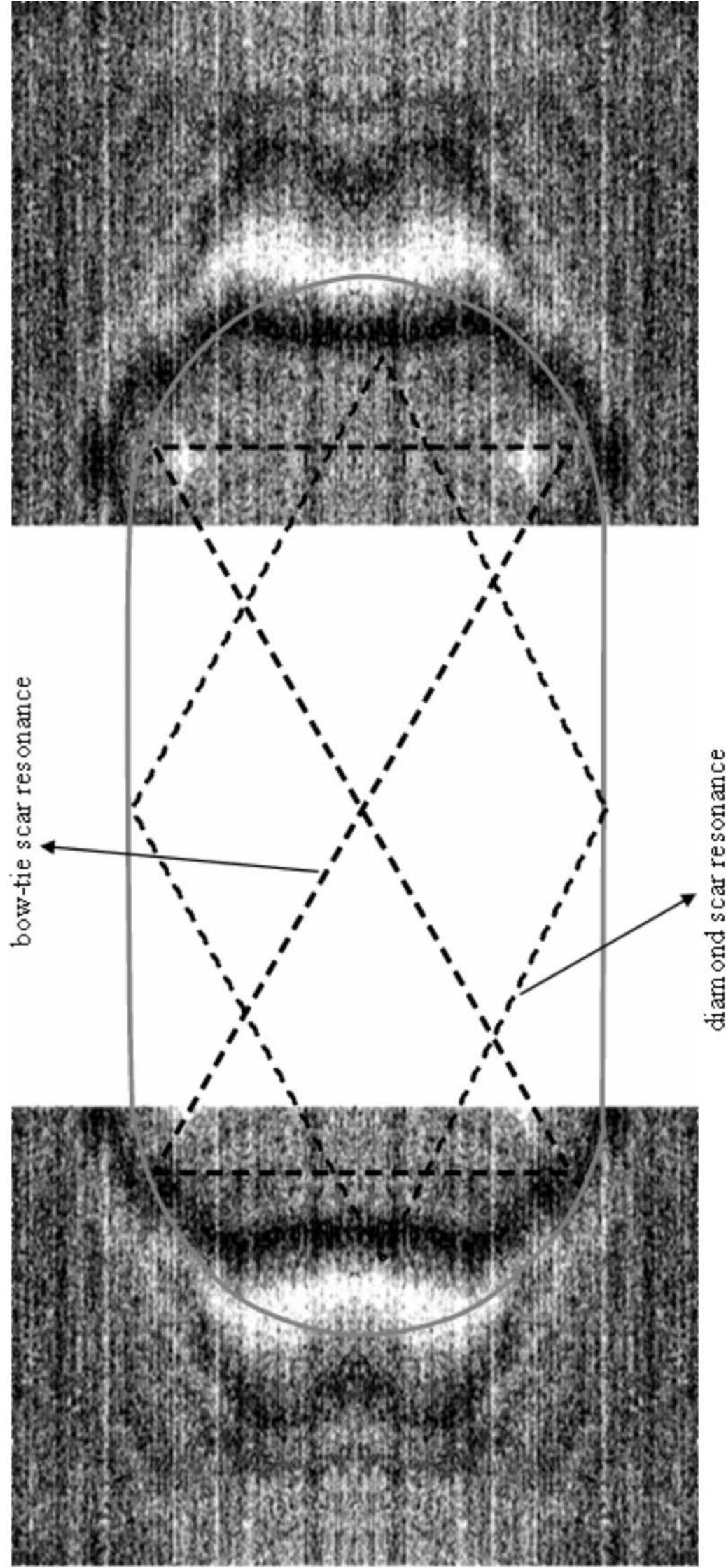

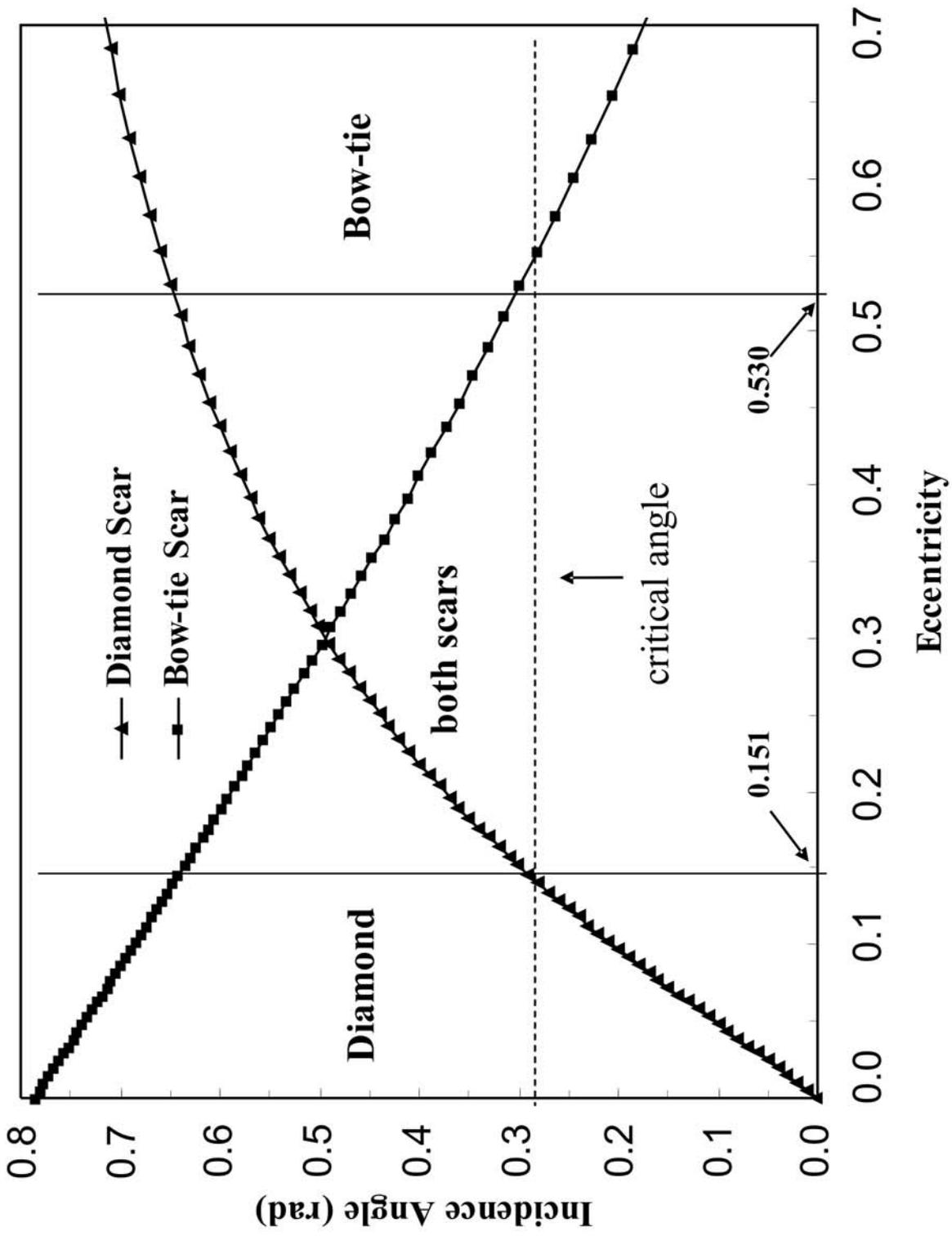